\documentclass[a4paper,11pt]{article}
\usepackage{pos}

\renewcommand{\logo}{\relax}
\title{Theory overview on LLP searches at the LHC}
\author*{Rebeca Beltrán}

\affiliation{ Department of Physics, University of Cyprus,\\
  P.O. Box 20537, 1678 Nicosia, Cyprus}

\emailAdd{beltran-lloria.rebeca@ucy.ac.cy}

\abstract{In these proceedings, we discuss heavy neutral leptons (HNLs) as well-motivated long-lived particle (LLP) candidates to illustrate the rich phenomenology that LLP searches can probe at the LHC. We briefly discuss the minimal HNL scenario, in which the phenomenology is governed by active-sterile neutrino mixing, and its characteristic displaced signatures. We then consider more general BSM scenarios with additional HNL interactions, showing how these can modify the main production and decay mechanisms, opening up new signatures beyond those of the minimal benchmark.}

\FullConference{14th Edition of the Large Hadron Collider Physics (LHCP2026)\\
18-22 May 2026\\
Paris, France\\}

\begin{document}
\maketitle

\section{Introduction}
Long-lived particles (LLPs) arise in a wide range of beyond the Standard Model (BSM) extensions \cite{Alimena:2019zri}, from supersymmetry to dark sector models or neutrino mass models. New particles with macroscopic lifetimes emerge because their decay widths are suppressed, for instance due to feeble interactions, a large hierarchy of mass scales, or phase space suppression. As a consequence, LLPs can travel macroscopic distances before decaying, leaving distinctive signatures at detectors. 

At the LHC, LLPs have received increasing attention over the last decade, leading to dedicated search strategies at the main detectors, as well as proposals for far detectors targeting LLPs \cite{MATHUSLA:2025zyt, ANUBIS:2025sgg, Feng:2017uoz, Gligorov:2017nwh}. Despite the experimental challenges associated with LLP searches, they provide a complementary way to probe regions of the parameter space of BSM models that can be difficult to access with traditional prompt searches.

Typically, existing searches and sensitivity studies focus on simple benchmark LLP scenarios. However, more general BSM models can present much richer phenomenology, which should also be studied to ensure that potential BSM signals are not overlooked. In these proceedings, we use heavy neutral leptons (HNLs) as a case study and explore how LLP searches at the LHC are sensitive not only to the minimal HNL scenario, but also to HNLs embedded in broader BSM frameworks.

\section{Searches for Heavy Neutral Leptons: minimal scenario}
Heavy neutral leptons, also called sterile neutrinos, are among the simplest and most extensively studied LLP candidates connected to neutrino physics (see \textit{e.g.} \cite{Drewes:2013gca, Abdullahi:2022jlv} for general HNL reviews). They arise naturally in various neutrino mass models (classical type-I seesaw \cite{Minkowski:1977sc} or inverse seesaw \cite{Mohapatra:1986bd}), and they interact with SM particles primarily through small mixings with the active neutrinos. The relevant charged and neutral current interactions with the SM leptons can be written as
\begin{equation}
- \mathcal{L}_{N} =  \frac{g}{\sqrt{2}} \sum_\alpha U_{\alpha N} \left( \overline{\ell_\alpha} \gamma^\mu P_L N \right) W_\mu + \frac{g}{2 \cos{\theta_W}} \sum_{\alpha , i} U_{\alpha i}^L U_{\alpha N}^\ast \left(\overline{N} \gamma^\mu P_L \nu_i\right) Z_\mu  + \text{h.c.},
\label{eq:lag_HNL}
\end{equation}
where $\alpha = \{e, \mu, \tau \}$, $i=\{1,2,3\}$ for the SM generations and $U_{\alpha N}$ denotes the active-sterile neutrino mixing. We refer to HNLs described by Eq.~\eqref{eq:lag_HNL} as \textit{minimal HNLs}. In phenomenological analyses of the minimal benchmark, without committing to any underlying neutrino mass model, the HNL mass, $m_N$,\footnote{The HNL mass term could be either Dirac or Majorana. The choice affects the possible final states into which the HNL can decay, resulting in a twice as large decay width for the Majorana case.} and the three active-sterile mixing parameters, $\{ U_{eN}, U_{\mu N}, U_{\tau N}\}$, are usually treated as independent parameters, oftentimes assuming only one mixing flavor to be non-zero.

At the LHC, minimal HNLs with GeV-scale masses could be produced, for instance, through $W$ and $Z$ boson decays, and then subsequently decay into final states containing SM leptons and quarks. For sufficiently small mixing values, HNLs are expected to be long-lived enough to leave displaced signatures in a main LHC experiment or in a planned far detector. For instance, $m_N= 50$~GeV and $|U_{eN}|^2 = 10^{-12}$ lead to $(c\tau)_N\approx 1$~m, while for $m_N= 500$~GeV and $|U_{eN}|^2 = 10^{-15}$ the proper decay length is $(c\tau)_N\approx 1$~mm.\footnote{These are obtained using the relation $c\tau \propto \Gamma^{-1}$, and the analytical decay width formulas from Ref.~\cite{Atre:2009rg}.}
The characteristic signature of minimal long-lived HNLs is typically a displaced vertex containing the HNL decay products, which can also be accompanied by a prompt charged lepton, depending on the HNL production process. LLP searches at the LHC targeting this signature have been able to probe minimal HNLs.

\begin{figure}[t]
\centering
\includegraphics[width=0.9\linewidth]{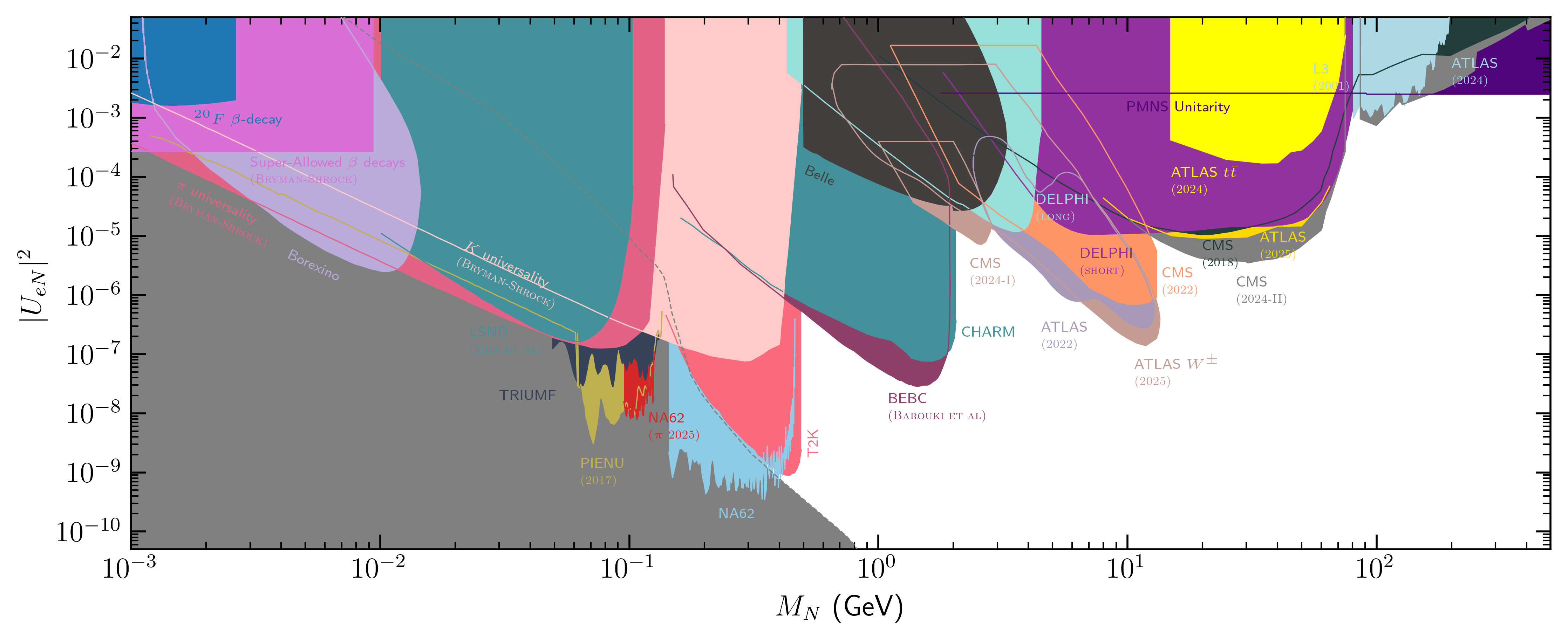}
\caption{Compilation of experimental bounds from Ref.~\cite{Fernandez-Martinez:2023phj} on the parameter space of minimal HNLs. It assumes a single Majorana HNL mixing only with the electron neutrino flavor. Analogous plots for $U_{\mu N}$ and $U_{\tau N}$ dominance are available on \href{https://github.com/mhostert/Heavy-Neutrino-Limits}{GitHub}, along with the references to all the experimental searches.}
\label{fig:bounds}
\end{figure}

In Fig.~\ref{fig:bounds}, we show the current experimental bounds on the minimal HNL parameter space, taken from Ref.~\cite{Fernandez-Martinez:2023phj}. The LLP searches at the LHC are particularly sensitive to HNL masses in the $  3-100 $~GeV range \cite{ATLAS:2025uah, ATLAS:2024fcs, CMS:2024ake, CMS:2024xdq}. They complement prompt searches \cite{ATLAS:2024rzi}, which can probe larger mixings for the same mass range, as well as other lower-energy experiments \cite{CHARM:1985nku, NA62:2020mcv, T2K:2019jwa} that predominantly probe lighter HNLs (with masses below 1 GeV). Far detectors proposed at the LHC, such as MATHUSLA \cite{MATHUSLA:2025zyt}, ANUBIS \cite{ANUBIS:2025sgg}, CODEX-b \cite{Gligorov:2017nwh} or MAPP \cite{Pinfold:2019nqj, Pinfold:2019zwp}, could further extend the sensitivity to smaller mixings in the GeV mass range and access parameter regions beyond the reach of the main detectors \cite{Wang:2025esc}, since they could probe HNLs with longer lifetimes.

Although the minimal HNL scenario provides a useful benchmark model for experimental searches, it does not encompass the full landscape of possible HNL phenomenology. If HNLs are only one piece of a larger BSM sector, as occurs in some neutrino mass models, additional interactions could substantially modify their phenomenology, leading to different signatures and changing the expected experimental sensitivity to HNLs. In the following section, we explore several examples of such beyond-minimal scenarios.

\section{Beyond minimal Heavy Neutral Leptons}
In a broader BSM context, additional light or heavy states could be present, and new interactions involving HNLs could modify both their production and decay. In particular, the couplings governing HNL production need not be the same as those controlling its decay and lifetime, which can lead to qualitatively different experimental signatures.

A convenient framework to describe such additional interactions is Effective Field Theory (EFT), in which higher-dimensional operators  containing the light degrees of freedom parametrize the low-energy effects of new physics at a presumably very high energy scale $\Lambda$,
\begin{equation}
\mathcal{L}_{\mathrm{EFT}}
= \mathcal{L}^{(4)}
+ \sum_{d\geq 5}\frac{1}{\Lambda^{d-4}}
\sum_i c_i^{(d)}\mathcal{O}_i^{(d)} \,.
\end{equation}
If the sterile neutrinos (or HNLs) have masses at the electroweak scale or below, they should be included among the low-energy degrees of freedom of the EFT, and the resulting frameworks are known as the $N_R$SMEFT \cite{Liao:2016qyd, Li:2021tsq} and $N_R$LEFT \cite{Li:2020lba, Chala:2020vqp}. The phenomenology of HNLs in these EFTs has been thoroughly studied in the past few years \cite{DeVries:2020jbs, Cottin:2021lzz, Beltran:2021hpq, Beltran:2022ast,  Beltran:2025ilg}. Here, we illustrate the impact of such effective interactions in the HNL phenomenology through a few representative scenarios.

\subsection*{HNLs with sizable magnetic moments}
\noindent
The magnetic properties of HNLs can be described by the dimension-5 dipole operators \cite{delAguila:2008ir, Aparici:2009fh}
\begin{equation} \mathcal{O}_{NN\gamma}^{ij} = \left( \bar{N}_{Ri}^{c} \sigma_{\mu \nu} N_{Rj} \right) F^{\mu\nu}\,, \qquad \mathcal{O}_{\nu N\gamma}^{\alpha i} = \left( \bar{\nu}_{L\alpha} \sigma_{\mu \nu} N_{Ri} \right) F^{\mu\nu}\,,
\end{equation}
where the first requires at least two HNLs and $i\neq j$. These operators induce the radiative decays $N_j\rightarrow N_i\gamma$ and $N_i\rightarrow \nu\gamma$, respectively. Even in the presence of active-sterile neutrino mixing, radiative decays can dominate the HNL decay width for sufficiently large dipole couplings, while one of the HNLs can remain long-lived. This gives rise to signatures featuring displaced or non-pointing photons, instead of the displaced vertices with jets or leptons characteristic of the minimal scenario \cite{Beltran:2024twr}. These exotic LLP signals can be searched for in the electromagnetic calorimeters of the LHC main detectors \cite{ATLAS:2022vhr, CMS:2012idp}.

\subsection*{Top-philic HNLs}
\noindent
If HNLs couple to top quarks through four-fermion operators, such as 
\begin{equation}
\mathcal{O}_{uN}^{13} = \left( \overline{u_R} \gamma_\mu t_R \right)\left(  \bar{N}_R \gamma^\mu N_R \right)\,, \qquad \mathcal{O}_{duNe}^{33} = \left( \overline{b_R} \gamma_\mu t_R \right)\left(  \bar{N}_R \gamma^\mu e_R \right) \,,
\end{equation}
motivated by possible top-philic new physics at high energy scales, their production at the LHC can proceed through rare top decays. Given the large top quark production rate at the LHC, this provides a complementary production mechanism to the electroweak channels of minimal HNLs. Moreover, if only four-fermion interactions with a pair of HNLs (\textit{e.g.}~$\mathcal{O}_{uN}^{13}$) are present, which cannot mediate the HNL decay, its lifetime can still be controlled by a small active-sterile mixing, allowing the HNL to decay with a macroscopic displacement from the main interaction point. In this scenario, the resulting signatures therefore combine numerous $b$-jets, associated with the top quark decays, with displaced vertices from the HNL decays. Sensitivity studies show that LLP searches at the LHC targeting these signatures could probe these top-philic HNLs \cite{Beltran:2025ilg}.

\subsection*{HNLs and axion-like particles}
\noindent
HNLs could also interact with other light BSM particles, such as axion-like particles (ALPs) \cite{deGiorgi:2022oks, Marcos:2024yfm}. If ALPs couple to gluons and HNLs through the dimension-5 operators
\begin{equation}
    \mathcal{O}_{NNa} = \partial_\mu a (\bar{N}_R \gamma^\mu N_R) \,, \qquad \mathcal{O}_{G\tilde{G}a} = a \ G_{\mu\nu}^A \tilde{G}^{\mu\nu, A}\,,
\end{equation}
with sizable couplings, ALPs could be abundantly produced at the LHC and subsequently decay into a pair of HNLs via $pp\to a^{(*)}\to NN$. The HNLs can nevertheless remain long-lived if their decays are still suppressed by small active-sterile mixing. In this case, the coupling responsible for HNL production is independent of the coupling controlling its decay and lifetime. This decoupling allows HNLs to be produced at significant rates without affecting the macroscopic decay length, ultimately modifying the parameter space accessible to LLP searches.

\begin{figure}[t]
\centering
\includegraphics[width=0.49\linewidth]{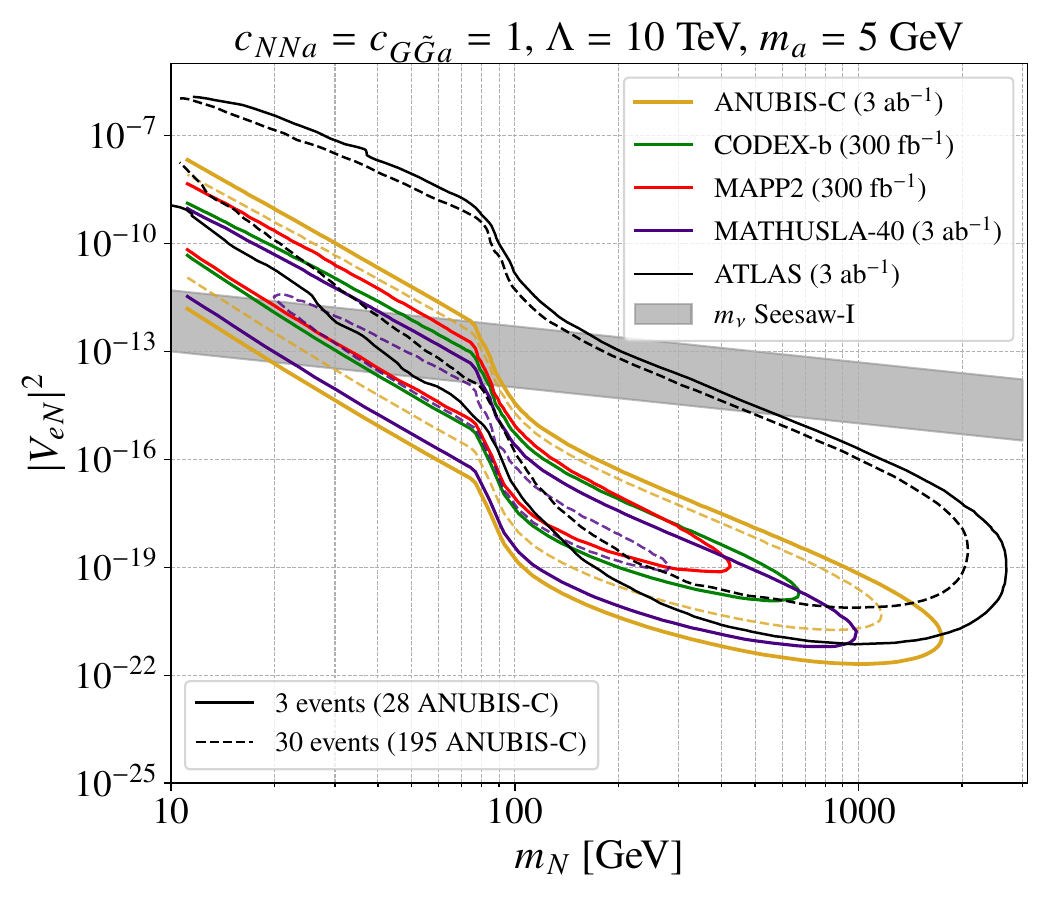}
\includegraphics[width=0.49\linewidth]{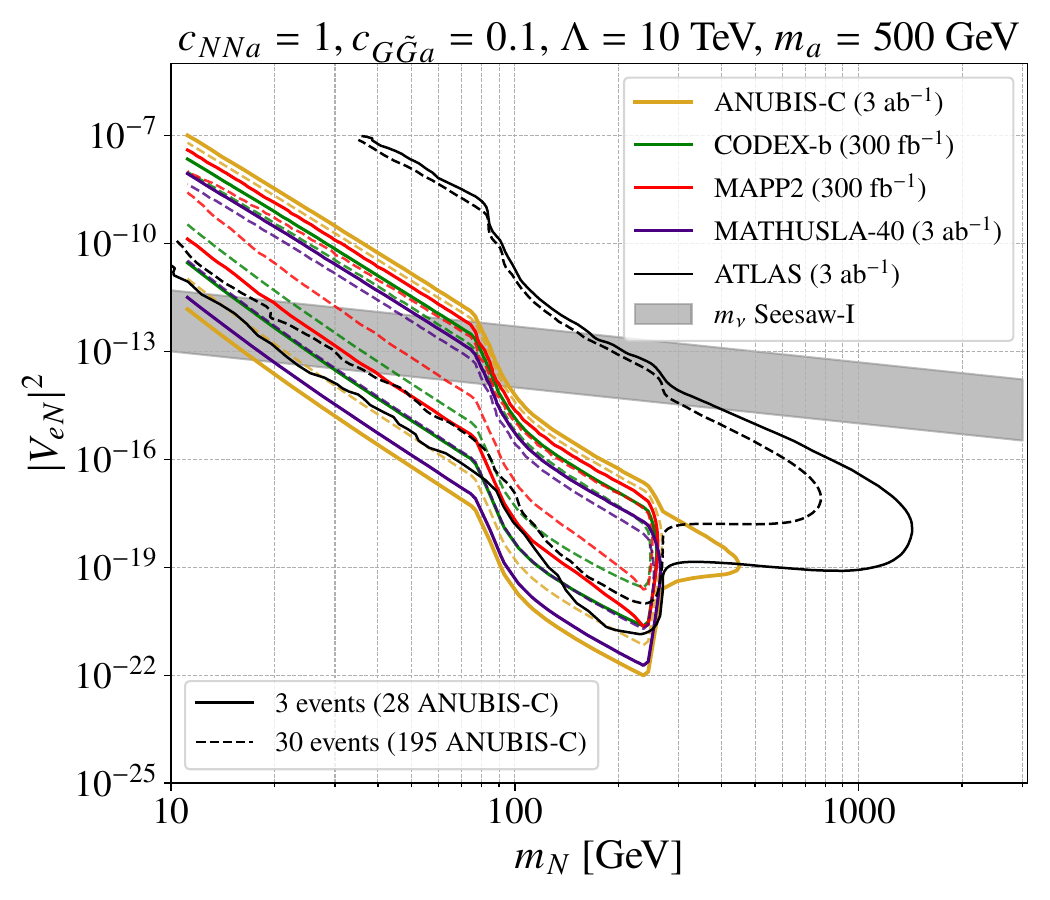}
\caption{Sensitivity reach in the $\{ m_N, |U_{eN}|^2\}$ plane for Majorana HNLs produced via the ALP portal \cite{Beltran:2025oqj}. Different colors correspond to different LLP searches at ATLAS and proposed far detectors at the LHC. The left (right) panel has a fixed ALP mass of $m_a= 5 \, (500)$~GeV, and considers equal (different) couplings for the two relevant operators. The new physics scale of the EFT is fixed to $\Lambda = 10$~TeV.  }
\label{fig:sensitivity}
\end{figure}

This is illustrated in Fig.~\ref{fig:sensitivity}, where we show the sensitivity prospects from Ref.~\cite{Beltran:2025oqj} for the high-luminosity LHC and some proposed far detector experiments in the $\{m_N, |U_{eN}|^2\}$ plane when $\mathcal{O}_{NNa}$ and $\mathcal{O}_{GGa}$ are switched on for two representative ALP masses. Compared with the minimal-HNL bounds shown in Fig.~\ref{fig:bounds}, the additional production mechanism allows LLP searches to probe regions of the HNL parameter space with much smaller active-sterile mixing. This clearly exemplifies how the non-minimal interactions have an impact on phenomenology and potentially extend the experimental reach of HNL searches.

\section{Conclusions}
Long-lived particle searches provide a powerful complementary avenue to explore weakly coupled physics beyond the SM at the LHC. Heavy neutral leptons provide a particularly well-motivated example, with displaced searches probing regions of the minimal parameter space not easily accessible for prompt searches. However, the minimal mixing scenario represents only one possible realization of HNL phenomenology. Additional interactions, which can be systematically described within effective field theories, can introduce new production and decay mechanisms and substantially modify the resulting LLP signatures and experimental reach. Exploring these extended scenarios is therefore essential to fully exploit the discovery potential of the LHC programme.

\vspace*{0.8cm}
\textbf{Acknowledgments. }
I thank the LHCP2026 organizers and conveners of the BSM-2 session for inviting me to give this talk. I also thank the co-authors of all the works presented here. I acknowledge support from the European Research Council (ERC) under the European Union’s Horizon 2020 research and innovation programme (Grant agreement No. 949451).

\bibliographystyle{JHEP}
\bibliography{Refs.bib}

\end{document}